\documentclass[aps]{revtex4}

\begin{document}
\draft

%
%

\preprint{Nisho-06/2}
\title{Light Cone Quantization and 
Savvidy Instability \\
in Dense Quark Matter}
\author{Aiichi Iwazaki}
\address{Department of Physics, Nishogakusha University, Ohi Kashiwa Chiba
  277-8585,\ Japan.} 
\date{Oct. 5, 2006}
\begin{abstract}
Solving instability of Savvidy vacuum in QCD
is a longstanding problem. Using light cone quantization
we analyze the problem not in the 
real confining vacuum but in dense quark matter
where gluons interact weakly with each other. 
We find a stable ferromagnetic ground state of gluons which carry 
a single longitudinal momentum.
Their states are composed as if they are confined in 
a two dimensional quantum well.
This supports our previous result
that gluons form a quantum Hall state in dense quark matter. 
\end{abstract}	
\hspace*{0.3cm}
\pacs{12.38.Lg, 12.38.-t, 12.38.AW, 24.85.+p, 73.43.-f \\
Light Cone Quantization, Quark Matter, Quantum Hall State}
\hspace*{1cm}

\maketitle

About 30 years ago, Savvidy\cite{savvidy} showed that
a color magnetic field is generated spontaneously
in the Yang-Mills gauge theory. Namely, when one calculates an effective potential 
of the color magnetic field using the one loop approximation, 
it is found that
the nontrivial color magnetic field is generated spontaneously.  
But, soon after it has been shown\cite{unstable} that some of gluons have imaginary energies
under the color magnetic field and produce an imaginary part in the 
effective potential.
The existence of the nontrivial imaginary part in the effective potential
implies the instability of
the vacuum with the color magnetic field.
Some of gluons are unstable in the vacuum.
We call it as Savvidy instability.
This magnetic instability of the vacuum in the Yang-Mills gauge theory
was expected by many authors to lead to a confining vacuum.
Namely, it was expected that the confining vacuum would be realized by
the condensation of the unstable gluons.
The subsequent analysis\cite{nielsen} of the gluons
has revealed the complication of the color magnetic flux
due to the production of additional magnetic field generated by
the unstable gluons.  
Although the formation of a lattice of the flux tube has been
argued\cite{nielsen},
any other clear pictures of 
such complicate states formed by the unstable gluons have not been presented.
Eventually, 
a confining vacuum could not be obtained.  

We have recently investigated the Savvidy instability in dense quark matter 
and shown\cite{iwa} that the instability is solved by 
the formation of a 
stable quantum Hall state\cite{qh,EI} of the unstable gluons.
Since perturbative arguments such as loop expansions 
are applicable in sufficiently dense quark matter, 
it is reliable that the spontaneous generation
of the color magnetic field arises in the matter. Although the quantum Hall states
is realized nonperturbatively due to the effect of gluon's repulsive self interactions,
this formation mechanism is well established in the physics of quantum Hall states of electrons.
This is similar to the formation mechanism of BCS states; BCS states arise due to the effect 
of attractive forces between electrons on the Fermi surface, even if the forces are fairly weak.
Hence, it is also reliable that the quantum Hall states of the gluons arise in the dense quark matter.
Consequently, we may understand that Savvidy instability is solved in the dense quark matter.
It is composed of quarks, the color magnetic field and
the colored quantum Hall state of gluons.
The phase of the quark matter is called as color ferromagnetic phase.
Although quarks occupy Landau levels, they do not 
form quantum Hall states in general.
( We have shown\cite{iwa} that the color ferromagnetic phase is realized in
the quark matter with lower densities than ones with which
color superconductivity\cite{color} is realized. 
Thus, the phase is phenomenologically more important than 
the color superconducting phase. 
We have discussed an astrophysical implication of the phase\cite{magnetar}
and also have pointed out the similarity\cite{cgc} between the gluons in the dense quark matter and
color glass condensate in nucleons. ) 

Quantum Hall states arise only in two 
dimensional space. For example,
quantum Hall states of electrons are realized in quantum wells of semiconductors,
which are effectively two dimensional.
Excitations with nontrivial momenta
perpendicular to the two dimensional well are forbidden
energetically as far as we are concerned with smaller energies ( or lower temperature )
than a finite gap. Thus,
only excitations with smaller energies than the gap are allowed
and they are excitations in the two dimensional well.
This is a feature of the two dimensional quantum well.  
Then, it is natural to ask
how two dimensional quantum Hall states of gluons
are formed in the 
three dimensional dense quark matter.
In this paper we analyze the problem as well as Savvidy instability,
using the light cone quantization\cite{light,cone}.
Since QCD Hamiltonian with color magnetic field can be well defined in the quantization, 
it is easy to analyze 
ground states of the gluons with the use of 
an approximation valid at small couplings.
As a result we find that the gluons in the lowest Landau level
form a ground state in which all of the gluons
have a single longitudinal momentum. 
Furthermore, the energies of excitations with the 
same longitudinal momentum as the momentum of the gluons in the ground state, 
are much smaller than the energies of excitations with nontrivial longitudinal momenta.
Such excitations arise in the two dimensional transverse space.
Thus, the gluons are two dimensional since 
their excitations are only allowed in the transverse directions
as far as we are concerned with much small energies. 
In this way two dimensional gluonic states arise effectively
in the three dimensional quark matter.
All of them carry the single 
longitudinal momentum. These gluons may form quantum Hall states, but 
we do not discuss in this paper how the two dimensional gluons form a quantum Hall state.
Hereafter we analyze SU(2) gauge theory.

First of all, we will give a brief review of 
our previous result\cite{iwa,EI} obtained with the use of equal-time quantization, that is,
how the two dimensional states of gluons arise 
in a ground state.
We assume the presence of a color magnetic field, $B_i^a\propto \delta_{i,3}\delta^{a,3}$, 
generated spontaneously; indices of $a$ and $i$ denote a coordinate of color and space, respectively.
Then, some of gluons under a spatially uniform color magnetic field, 
$B=\epsilon_{3ij}\partial_i A_j^3$,
have imaginary energies such that $E^2=k^2-gB$.
These gluons ( $A_l^1+iA_l^2\propto e^{-iEt}$ perpendicular to
magnetic field, $B$, in color space ) occupy the lowest Landau level
and have color magnetic moments parallel to $B$.
Here $g$ is gauge coupling constant and $k$ is a momentum parallel to
the magnetic field $\vec{B}=(0,0,B)$.
Indices, $j$ of $A^{j}_l$ denote colors of SU(2).  
The other stable gluons occupy higher Landau levels and their energies
are given by 
$E^2=k^2+gBn$ with integer $n \geq 0$.
Thus, the unstable gluons with small momentum such as $k^2<gB$, have imaginary
energies and their amplitudes grow up rapidly in time. Among them,
the most unstable gluons are the ones with vanishing momentum $k=0$.
( There are infinitely many degenerate states with $k=0$ in the lowest Landau level. )  
Hence, their wave functions are uniform in the direction
parallel to the magnetic field. These most unstable gluons with $k=0$ 
are expected to form a new stable ground state similar to the case in Higgs model
with a negative mass term $-m^2|\phi|^2$. 
In the model,
unstable Higgs modes with energies such as $E^2=\vec{k}^2-m^2 < 0$, are present 
in a naive vacuum $\langle \phi \rangle=0$. Among them, 
the most unstable modes with $\vec{k}=0$
form a new ground state, $\langle \phi \rangle \neq 0$ with the 
condensation of the field $\phi $.
Similarly, we expect that 
the gluons only with $k=0$ form a stable ground state. 
They are two dimensional objects since
they are uniform in the direction of $\vec{B}$. 
In this way the two dimensionality 
of gluonic ground state arises
since the most unstable gluons only with $k=0$ 
make a stable ground state. It is reasonable to assume that any gluons with $k\neq 0$ do not
contribute to the formation of the ground state. This is because
the similar phenomena arise in the Higgs model. 
This is our previous result derived with the use of the equal-time quantization.
( It is well known that bosons as well as fermions
in two dimensional space can form
a quantum Hall state if repulsive interaction between them is present. Hence,
the two dimensional gluons may make a quantum Hall state 
due to the repulsive self-interaction\cite{qh,nakajima}. )


In such a circumstance, 
we wish to examine in detail whether ground state of gluons
are two dimensional or not, by using the light cone quantization.
In the formulation, 
the states with imaginary light cone energy never appear.
Instead, a naive Fock vacuum is not the lowest energy state.
The real ground state with the lowest energy is formed as a composite 
state of gluons in the lowest Landau level.
In this paper we do not investigate the real confining vacuum which
is a complicated composite of strongly interacting gluons; the problem of
obtaining such 
a real QCD vacuum is beyond our scope\cite{kondo}.
Instead, we investigate a ground state of gluons 
in dense quark matter, in which gluons are weakly interacting with each other.  
We assume the presence of
a color magnetic field, which is generated spontaneously.
Here, we do not address a question of spontaneous generation of color 
magnetic field in the light cone quantization.

Now, we show that a ground state of gluons is two dimensional,
that is, it depends trivially on a longitudinal momentum $p^+$. 
We use the notations of the light cone time coordinate, $x^+=(x^0+x^3)/\sqrt{2}$ and longitudinal
coordinate, $x^-=(x^0-x^3)/\sqrt{2}$. Transverse coordinates are denoted by $x^i$,
or $\vec{x}$.
We assume a finite length, $-L\leq x^- \leq L$,
in the longitudinal space and impose a periodic boundary condition such that
$A^a_j(x^{-}=L)=A^a_j(x^{-}=-L)$. Then,
corresponding momentum becomes discrete denoted by $p_n^+=n\pi/L$ with integer $n$.
Light cone components of gauge fields, $A^+,\,A^-$, $A_i$, are defined similarly.
Then, the Hamiltonian, $H$, with the light cone gauge, $A^+=0$ is given by

\begin{equation}
\label{H}
H = \frac{1}{4} F_{ij}^a\, F_{ij}^a + 
\frac{g^2}{2}\rho^a\, \frac{1}{(-\partial_-^2)}\,\rho^a ,
\end{equation}
with field strength, $F_{ij}^a=\partial_iA^a_j-\partial_jA^a_i+g\epsilon_{abc}A^{b}_iA^{c}_j$,	
where color indices $a$ run from $1$ to $3$ and space indices, $i,j$ run from $1$ to $2$.
$\rho^a$ is defined by

\begin{equation}
\rho^a\equiv (D_iA_i)^a=(\partial_i\delta^{ab}+g\epsilon^{a3b}A_i^B)A_i^b
\end{equation}
where $A_i^B$ denotes the gauge potential of color magnetic field, 
which is assumed to direct into $\sigma_3$ in color SU(2);
$B=\partial_1A_2^B-\partial_2A_1^B$.
We have neglected a dynamical gauge potential, $A_i^{a=3}$ aside from the classical 
one, $A_i^B$ since it does not couple directly with $A_i^B$. We have only taken 
dynamical gauge fields, $A_i^{a=1,2}$ perpendicular in color space to the color magnetic field.
They form Landau levels under $B$.
We treat only the quantum effects of the gluons, but treat quarks classically for simplicity.
Their color charge neutralize the color charge
of the gluons.

We make a comment that our treatment of "zero mode"\cite{yamawaki,yamawaki2}
 in the light cone quantization
is similar to the one used by Thorn\cite{thorn}: 
We quantize gauge fields 
in the finite volume, $-L\leq x^{-}\leq L$, and neglect zero modes of the fields.
Consequently, Hamiltonian becomes a simple form involving
at most quartic terms of creation or annihilation operators
in addition to quadratic ones. As has been shown\cite{thorn} in a two dimensional model
of scalar field, the true ground state can be gripped even if
we neglect the zero mode of the field, at least in the limit of $L\to \infty$.  
We assume that it also holds in the gauge theory.
We may justify 
neglecting the zero mode in the analysis of dense quark matter as follows.
That is, our concern is not the real vacuum, but a ground state of gluons
in dense quark matter. 
The zero mode may play an important role in the real vacuum of strongly interacting gluons.
But, it may not play such a role 
in a ground state of gluons weakly interacting with each other in the dense quark matter. 
Typical energy scale of quarks and gluons in the quark matter 
is given by the chemical potential of the quarks and is much larger than $\Lambda_{QCD}$.
In such dense quark matter the zero mode does not play an important role for
realizing the ground state.
It is similar to the case of QCD at high energy scattering\cite{qcd}
where the typical energy scale is much higher than $\Lambda_{QCD}$. Hence,
the zero mode does not play an important role 
for realizing so-called color glass condensation.
Therefore, it is reasonable to neglect the zero modes 
of the gluons in the quark matter.

We assume the spontaneous generation of the color magnetic field, $B$,
in the light cone quantization.
The condition of neglecting the zero mode requires the coherent length of $B$ in $x^-$
being shorter than $L$, but becoming infinite as $L\to \infty$.
Thus, our discussion below is limited for the case such that 
momentum scale, $p^+$ of gluons is larger than the inverse of the coherent length
of $B$.

In the light cone gauge only dynamical variables are transverse components, $A_i^a$ of gauge fields.
This can be expressed in terms of creation and annihilation operators,

\begin{equation}
A_i^b=\sqrt{\frac{\pi}{L}}\sum_{p^+>0}\frac{1}{\sqrt{2\pi p^+}}
(a_{i,p}^b(x^+,\vec{x})e^{-ip^+x^-}+ a_{i,p}^{b\dagger}(x^+,\vec{x})
e^{ip^+x^-}),
\end{equation}
with $p^+ =\pi n/L$ ( integer, $n\geq 1$ ),
where operators, $a_{i,p}^l$, satisfy the commutation relations,
$[a_{i,p}^b(x^+,\vec{x}),
a_{j,k}^{\dagger c}(x^+,\vec{y})]=\delta_{ij}\delta^{bc}\delta_{pk} \delta(\vec{x}-\vec{y})$,
with other commutation relations being trivial. 
As we have mentioned before,
we have neglected the zero modes, $p^+=0$, of the gauge fields. 
  
Then, the gauge fields satisfy the equal time, $x^+$, commutation relation,

\begin{equation}
[\partial_-A_i^a(x^+,x^-,\vec{x}),A_j^b(x^+,y^-,\vec{y})]=
-i\delta_{ij}\delta^{ab}\delta(\vec{x}-\vec{y})\Bigl(\delta(x^- -y^-)-\frac{1}{2L}\Bigr),
\end{equation}
where the last factor, $1/2L$, in the right hand side of the equation 
comes from neglecting the zero modes
of the gauge fields.

We should mention that the second term in $H$ represent a Coulomb interaction. 
It is derived by solving a constraint equation, $\partial_-^2 A^{-,a}=\rho^a$,
that is, Gauss law associated with the light cone gauge condition, $A^+=0$.
In order to assure that the gauge field, $A^-$ is periodic in $x^-$,
the zero mode of $\rho$ ( $\rho\propto \sum_{n=\mbox{integer}}\rho_n\, e^{i\pi nx^-/L}$ )
must vanish; $\rho_{n=0}=0$. Then, the operation of $1/(-\partial_-^2)$ is 
well defined. The condition of $\rho_{n=0}=0$ implies that the total color charge, 
$\int_{-L}^{L} dx^- \rho_{tot}$,
vanishes. Here we should include background classical color charge of quarks in $\rho_{tot}$.
$\rho$ in eq(\ref{H}) should be replaced by $\rho_{tot}=\rho-\rho_{quark}$
with $\rho_{tot,n=0}=0$.
This requirement of $\rho_{tot,n=0}=0$, is consistent with the condition, $A^-_{n=0}=0$.

We now rewrite the Hamiltonian in terms of "charged vector fields", $\Phi_i=(A_i^1+iA_i^2)/\sqrt{2}$,
which may be decomposed into spin parallel ( anti-parallel ) component, $\Phi_p=(\Phi_1+\Phi_2)/\sqrt{2}$
( $\Phi_{ap}=( \Phi_1-\Phi_2)/\sqrt{2}$ ).
These fields transform as Abelian charged fields under the $U(1)$ gauge transformation, 
$A_i\to U^{\dagger}A_iU$ with $U=exp(i\theta \sigma_3)$.

Then, using the fields, $\Phi_{p}$ ( $\Phi_{ap}$ ), we obtain the following Hamiltonian,

\begin{eqnarray}
\label{H'}
H=\frac{1}{2}B^2&+&\Phi_{p}^{\dagger}(-\vec{D}^2-2gB)\Phi_{p} +\Phi_{ap}^{\dagger}(-\vec{D}^2+2gB)\Phi_{ap} \\ \nonumber
&+&\frac{g^2}{2}(|\Phi_{p}|^2-|\Phi_{ap}|^2)^2+\frac{g^2}{2}\rho\,\frac{1}{(-\partial_-^2)}\,\rho\,\,\,, 
\end{eqnarray}
with $\vec{D}=\vec{\partial}+ig\vec{A^B}$,
where $\rho$ is given by

\begin{equation}
\rho=i(\Phi_{p}^{\dagger}\partial_-\Phi_{p} -\partial_-\Phi_{p}^{\dagger}\Phi_{p} +
\Phi_{ap}^{\dagger}\partial_-\Phi_{ap} -\partial_-\Phi_{ap}^{\dagger}\Phi_{ap})-\rho_{quark}\,\,.
\end{equation}

The first term in eq(\ref{H'}) represents the classical energy of the color magnetic field, and second ( third ) term
does the kinetic energy of the charged gluons with spin parallel ( anti-parallel ) under the color
magnetic field, $B$. The forth term represents the energy of the repulsive self-interactions. The last term
represents the Coulomb energy coming from the second term with $\rho^{a=3}$ component in eq(\ref{H}).

Since eigenstates of the operator $\vec{D}^2$ are classified by Landau levels,
the second and the third terms can be rewritten as,

\begin{equation}
\sum_{n=0,1,2,,,}\Big(\Phi_{p,\,n}^{\dagger}(2n-1)gB\Phi_{p,\,n}+\Phi_{ap,\,n}^{\dagger}(2n+3)gB\Phi_{ap,\,n}\Big),
\end{equation}
where the fields, $\Phi_{p,\,n}$ ( $\Phi_{ap,\,n}$ ) denote operators in the Landau level specified by integer $n$.
( We have implicitly assumed integration over the transverse directions in the above equation. )

Now, we take only the field, $\Phi_{p,\,n=0}$ in the lowest Landau level, $n=0$,
that is, the component having negative kinetic energy. 
It is most important, among others, for realizing the ground state
of the Hamiltonian in the limit of strong magnetic field, $B$.
The field corresponds to the unstable gluon in our previous discussions\cite{iwa} with the use of
the time-like quantization.
Therefore, we obtain the following reduced Hamiltonian for analyzing the ground state of the system,
 
\begin{equation}
\label{hr}
H_r=-gB|\Phi|^2+\frac{g^2}{2}|\Phi|^4+\frac{g^2}{2}\rho_r\,\frac{1}{(-\partial_-^2)}\,\rho_r,
\end{equation}
with $\rho_r\equiv i(\Phi^{\dagger}\partial_-\Phi - \partial_-\Phi^{\dagger}\Phi) -\rho_{quark}$,
where we have put $\Phi\equiv \Phi_{p,\,n=0}$ for simplicity.
The field, $\Phi$, can be expressed by using creation and annihilation operators,

\begin{equation}
\Phi=\sqrt{\frac{\pi}{L}}\sum_{p>0,m=0,1,2,,}\frac{1}{\sqrt{2\pi p}}(a_{p,m}\phi_m(\vec{x})e^{-ipx}+
b_{p,m}^{\dagger}\phi_m^{\dagger}(\vec{x})e^{ipx}),
\end{equation}
where simplified notation such as $x=x^-$ and $p=p^+$ is exploited and
will be used below.
$\phi_m(\vec{x})=g_mz^m exp(-|z|^2/4l^2)$ represents normalized
eigenfunction of $\vec{D}^2$ with angular momentum, $m$, in the lowest Landau level, 
$\int d^2\vec{x} \phi_m^{\dagger}\phi_n=\delta_{m,n}$;
$z=x_1+ix_2$ and
$g_m\equiv \frac{1}{\pi m!(2l^2)^{m+1}}$ with $l^2\equiv 1/gB$. 
$a_{p,m}$ and $b_{p,m}$ satisfy the commutation relations;
$[a_{p,m},a^{\dagger}_{k,n}]=\delta_{p,k}\delta_{m,n}, \quad [b_{p,m},b^{\dagger}_{k,n}]=\delta_{p,k}\delta_{m,n}, 
\quad \mbox{others}=0$.

When we express the first term in eq(\ref{hr}) in terms of the operators, $a_{p,m}$ and $b_{p,m}$,

\begin{equation}
\int_{-L}^{L}dx\,d^2\vec{x}:-gB|\Phi|^2:=-gB\sum_{p>0,m}\frac{1}{p}(a^{\dagger}_{p,m}a_{p,m}+b^{\dagger}_{p,m}b_{p,m}),
\end{equation}
we find that there exist states with lower energies than a trivial Fock vacuum, $|\mbox{vac}>$; 
$a_{p,m}|\mbox{vac}>=b_{p,m}|\mbox{vac}>=0$. Namely, gluons in the lowest Landau level
are produced spontaneously to form a state with lower energy than that of the vacuum.
The production of the gluons is limited by the second term in eq(\ref{hr}) representing 
repulsion among the gluons. 
This is similar to the case of Higgs model. Contrary to the model,
the gluons do not condense.
Thus, we postulate $ \langle \Phi \rangle= 0$.
( This is consistent with neglecting zero mode. )
In order to find the ground state of the gluons,
we will express approximately the energy, $\langle H_r \rangle $, of the state
in terms of the distribution functions,
$\langle a^{\dagger}_{p,m}a_{p,m}\rangle$ and $\langle b^{\dagger}_{p,m}b_{p,m}\rangle$.
Then, by minimizing the energy
we will find the momentum, $p$, and angular momentum, $m$, distribution of gluons. 
The state of the gluons may carry
non vanishing color charge density, in general. 
The fact that quarks and gluons are weakly interacting in 
the dense quark matter certificates that our approximation gives rise to
reliable results; any quantum corrections to the results are small.

Before finding the ground state of the Hamiltonian, $H_r$, we should note that
there are two conserved quantities such as total color charge and momentum,

\begin{eqnarray}
Q&=&\int_{-L}^{L}dx\,d^2\vec{x}\,\rho_r=\sum_{p>0,m}(a^{\dagger}_{p,m}a_{p,m}-b^{\dagger}_{p,m}b_{p,m})
-\int_{-L}^{L}dx\,d^2\vec{x}\rho_{quark}=0 \nonumber \\	
P_{total}&=&\sum_{p>0,m}p(a^{\dagger}_{p,m}a_{p,m}+b^{\dagger}_{p,m}b_{p,m}),
\end{eqnarray}  
where we have required that the total color charge of gluons and quarks must vanish.
The ground state must be found under the condition of 
the conserved quantities being given. 
Since our concern is the ground state of gluons in dense quark matter, not real vacuum,
the total color charge of the gluons can be nonzero and is neutralized by the color charge of quarks.

In order to evaluate the expectation value of $H_r$ with the ground state, $|g\rangle$,
we assume the following approximation,

\begin{eqnarray}
\label{ap}
\langle a_{\alpha}^{\dagger}a_{\beta}^{\dagger}a_{\gamma}a_{\delta}\rangle &\simeq& 
\langle a_{\alpha}^{\dagger}a_{\gamma}\rangle \langle a_{\beta}^{\dagger}a_{\delta}\rangle +
\langle a_{\alpha}^{\dagger}a_{\delta}\rangle \langle a_{\beta}^{\dagger}a_{\gamma}\rangle, 
\,\,\,\mbox{for}\,\alpha\neq \beta,\quad \mbox{similar for}
\,\,\,b_{\alpha} \label{12} \\
\langle a_{\alpha}^{\dagger}b_{\beta}^{\dagger}a_{\gamma}b_{\delta}\rangle &\simeq& 
\langle a_{\alpha}^{\dagger}a_{\gamma}\rangle \langle b_{\beta}^{\dagger}b_{\delta}\rangle,
\,
\langle a^{\dagger}a^{\dagger}b^{\dagger}a\rangle =\langle a^{\dagger}b^{\dagger}b^{\dagger}b\rangle =
\langle a^{\dagger}aab\rangle=\langle b^{\dagger}bba \rangle=0, \label{13} \\
\langle a_{\alpha}^{\dagger}a_{\beta}\rangle&\propto& \delta_{\alpha,\beta},\quad
\langle b_{\alpha}^{\dagger}b_{\beta}\rangle \propto \delta_{\alpha,\beta} \label{14}
\end{eqnarray}
where indices, $\alpha,\,\beta$, , , denote a set of $p$ and $m$.
Namely, we assume no mixing among states with different
$p$ and $m$ in eq(\ref{12}). Furthermore, 
we assume no pair creations and annihilations
in eq(\ref{13}) so that the numbers of particle and antiparticle
are conserved, respectively. 
We also postulate momentum and angular momentum conservation in eq(\ref{14}).
These formulae are satisfied by  
eigenstates of the number operators, $a_{\alpha}^{\dagger}a_{\alpha}$ and $b_{\alpha}^{\dagger}b_{\alpha}$.
Hence, we minimize the Hamiltonian by using the eigenstates of the number operators in our approximation.

Using the approximation, we evaluate an expectation value, $\langle :H_r: \rangle$, of 
normal ordered Hamiltonian, $:H_r:$ 
and find the ground state
minimizing the expectation value,

\begin{eqnarray} 
\label{energy}
\langle :H_r: \rangle =-&gB&\sum_{p>0,m}\Bigl(a(p,m)+b(p,m)\Bigr)+ \nonumber \\
&+&\frac{g^2}{2L}\sum_{p,q>0,m,n}
\Bigl(a(p,m)a(q,n)+b(p,m)b(q,n)+2a(p,m)b(q,n)\Bigr)N_{m,n} \nonumber \\
&+&\frac{g^2}{2L}\sum_{p\neq q>0}\biggl(\frac{p+q}{p-q}\biggr)^2\sum_{m,n}
\Bigl(a(p,m)a(q,n)+b(p,m)b(q,n)\Bigr)N_{m,n}\nonumber \\
&+&\frac{g^2}{2L}\sum_{p,q>0}\biggl(\frac{p-q}{p+q}\biggr)^2\sum_{m,n}2a(p,m)b(q,n)N_{m,n},
\end{eqnarray}
with $N_{m,n}=(m+n)!\,(\pi m!\,n!\,2^{m+n+2}\,l^2)^{-1}$,
where $a(p,m)$ and $b(p,m)$ are defined by

\begin{equation}
\langle a_{p,m}^{\dagger}a_{q,n}\rangle =\delta_{p,q}\delta_{m,n}\,p\, a(p,m), \quad
\langle b_{p,m}^{\dagger}b_{q,n}\rangle =\delta_{p,q}\delta_{m,n}\,p\, b(p,m),
\end{equation}  
where $a(p,m)$ and $b(p,m)$ represent distributions of $p$ and $m$ in the ground state, respectively.
We have used the color neutrality condition, $\rho_{r,n=0}=0$ in the derivation of eq(\ref{energy});
the condition is imposed as a constraint such as $p\neq q$ in the third term.

It is easy to see
that the expectation value of the color charge density becomes uniform in 
longitudinal direction, $x=x^-$,

\begin{equation}
\langle \rho_r \rangle =\frac{1}{L}\sum_{p>0,m}p\Bigl(a(p,m)-b(p,m)\Bigr)|\phi_m(\vec{x})|^2
-\rho_{quark},
\end{equation}
which may depend on transverse coordinate, $\vec{x}$.

In eq(\ref{energy}) the first term represents the kinetic energy in the Landau level
and the second term does the energy of the repulsion between gluons.
These two terms are denoted by $E_1$.
The third and forth terms represent the Coulomb energy between gluons 
and are denoted by
$E_2$.
Obviously, these terms are non negative except for the first one.
Therefore, we can find a ground state with the lowest energy, $\langle :H:\rangle\equiv E_1+E_2$,
by minimizing the first two terms, $E_1$ and the last two terms, $E_2$ of the Coulomb energy,
respectively. Simultaneously, we need to take into account the condition that
the state carries a given total momentum 
and vanishing total color charge.
We may assume that the color charge of quarks, $\rho_{quark}$ is spatially uniform.

It is easy to minimize the first two terms, i.e. $E_1$, which is given by a set of values,
$c(m)\equiv \sum_{p>0}(a(p,m)+b(p,m))$. This is because $E_1$ can be rewritten 
such that $E_1=-gB\sum_m c(m)+g^2/(2L)\sum_{m,n}c(m)H_{m,n}c(n)$.
Thus, minimizing the energy, $E_1$, does not
determine the dependence of $a(p,m)$ or $b(p,m)$ on the longitudinal momentum, $p=p^+$.
It simply gives the summation over the momentum, $p$, namely, $c(m)\propto gBL/g^2$. 
The dependence on $p$ is determined only by minimizing
the Coulomb energy, $E_2$. The energy, $E_2\geq 0$, can be minimized easily by assuming
that the ground state depends only on a single momentum, $p=p_0$, that is, $a(p,m)\propto \delta_{p,p_0}$
and $b(q,n)\propto \delta_{q,p_0}$. This distribution of the momentum 
leads to the minimum, $E_2=0$. 
( Any other distributions with the dependence on various momenta 
give rise to higher energies ( $>0$ ). )
It follows from the distribution that 
the color charge density is given such that 
$\langle \rho_r \rangle=p_0\sum_m\Bigl(a(p_0,m)-b(p_0,m)\Bigr)|\phi_m(\vec{x})|^2/L-\rho_{quark}$.
For the color charge density to vanish,
$a(p_0,m)-b(p_0,m)$ should be independent of $m$.
Then, the color charge density of the gluons becomes uniform in $\vec{x}$,
and can cancel that of quarks, $\rho_{quark}$ 
since $\sum_m |\phi_m(\vec{x})|^2=1/(2l^2\pi)$.
On the other hand,
the total momentum is given by 
$P_{tot}=\sum_mp_0^2(a(p_0,m)+b(p_0,m))=p_0^2\sum_m\,c(m)$. 

Therefore, the ground state minimizing the energy in eq(\ref{energy})
is characterized by the trivial distribution of the longitudinal momentum;
$\langle a_{p,m}^{\dagger}a_{p,m}\rangle \propto \delta_{p,p_0}$ and 
$\langle b_{p,m}^{\dagger}b_{p,m}\rangle \propto \delta_{p,p_0}$.
The Coulomb energy, $E_2$, of the state vanishes.
All of gluons in the lowest Landau level occupy the states with a single momentum,
e.g. $p=p_0$.
It implies that the ground state is composed of two dimensional gluons occupying
the lowest Landau level. 
It is also important to note that the displacement of a gluon with the momentum, $p_0$
in the ground state to
a state with a momentum $k\neq p_0$ gives rise to an energy gap, $\Delta E_k$;
$\langle k|:H_r:|k\rangle =\langle :H_r: \rangle + \Delta E_k$ 
( $\Delta E_k > 0$ for any $k$ )
where $|k\rangle \equiv a^{\dagger}_{k,m}a_{p_0,m}|g\rangle$.
We note that $\Delta_k\to \infty$ as $k\to p_0$.
This simple argument suggests the existence of a finite gap energy 
needed to excite modes with longitudinal
momenta different with $p_0$. On the other hand, we can show that there exist gapless excitations
with the momentum, $p_0$, which have different distributions in $m$ with 
$a(p_0,m)$ and $b(p_0,m)$ in the ground state. 
The fact implies that gluons are confined in a two dimensional quantum well 
extending in the transverse directions.
Only motions are allowed in the well
as far as we are concerned with smaller energies than the gap energy.
Motions in the longitudinal direction are visible
only when we are concerned with higher energies than the gap energy.
On this point, more detail analysis is necessary to demonstrate the conclusion.

Here, we wish to mention that for $P_{total}$ to be finite 
in the limit of $L\to \infty$, $p_0$ goes to $0$ 
in the limit because $c(m)\propto L$.
On the other hand,
the color charge density of gluons, 
$\propto p_0 \Big(a(p_0,m)-b(p_0,m)\Big)$, can 
take any finite value even in the limit. 
Hence, all of gluons occupy the states with a vanishingly small longitudinal
momentum. In the terminology used in high energy scattering of hadrons 
or nuclei, parton ( gluons ) distribution behaves such as $\delta(x)$ since
 $x=p_0/P_{tot}\to 0$. ( The distribution contains only contributions of
charged vector field, $\Phi_{\pm}$ or $A_i^{a=1,2}$.
If we take into account 
momentum distribution of the gauge field, $A_i^{a=3}$
neglected in our argument, it would give a parton distribution
with a finite support in $x$. )

We have not yet shown that the gluons forms quantum Hall states, but
have only shown that they form two dimensional states. It is a necessary
condition for the realization of quantum Hall states of the gluons. 
Whether or not the quantum Hall states of the gluons are made,
depends on the color charge density of the gluons.
If a specific condition of "filling factor", $\nu=2\pi\rho_rL/gB$
is satisfied,
the quantum Hall state can be realized, that is, 
$\nu=2\pi p_0\sum_m\Bigl(a(p_0,m)-b(p_0,m)\Bigr)|\phi_m(\vec{x})|^2/gB
=1/2,\,1/4,\,,\,,$ .
Otherwise, 
gluons simply form a two dimensional compressible state,
i.e. gapless state.

As we have shown,
the two dimensionality of the ground state of gluons arises from the Coulomb interaction,
$\rho\frac{1}{-\partial_-^2}\,\rho$
in the longitudinal direction, $x^-$. In our approximation,
the Coulomb energy, $E_2$ is a positive semi definite. Thus, 
the ground state is given by the state with $E_2=0$. 
When we apply the similar approximation to quarks coupled with the magnetic field, 
we obtain a negative semi definite Coulomb energy.
Thus, we do not obtain the two dimensionality in the quarks.
In the ground state of the quarks, 
the distribution of the longitudinal momentum does not take a form of a delta
function.
The result in the quarks or fermions is natural physically.
Electrons in metals never take a form of two dimensional gases
under external magnetic field
except for ones confined in two dimensional space, e.g. quantum wells.

The difference between the case of gluons and that of quarks comes from
their difference in statistics. In order to see this point more explicitly,
we take a following simple model of a Coulomb interaction,

\begin{equation}
V_c=\frac{g^2}{2}:\rho\,\frac{1}{(-\partial_-^2)}\,\rho:\quad \rho(x)=\psi^{\dagger}(x)\psi(x)
-\langle\psi^{\dagger}\psi\rangle
\end{equation} with $-L\leq x\leq L$,
where the field, $\psi$ represents boson or fermion with an appropriate boundary condition
at $x=\pm L$.


Suppose that we have two states with momentum, $p>0$ and $q( \neq p)>0$, whose wave functions 
are denoted as $f(p,x)$ and $f(q,x)\propto e^{-iqx}$. We define the field operator such that

\begin{equation}
\psi(x)=a_p\,f(p,x)+a_q\,f(q,x),
\end{equation}
with annihilation operators, $a_p$ and $a_q$ satisfying
commutation relations,
$[a_p,a_q^{\dagger}]_{\pm}=\delta_{p,q}$, ,  .
As in the previous case, we extract zero modes of $\rho$ 
for $1/(-\partial_-^2)$ to be well defined. Thus, 
$\psi^{\dagger}\psi=a_p^{\dagger}a_qf^{\dagger}(p,x)f(q,x)+h.c$.
When we evaluate the expectation
value of $V_c$ with the use of a state, $a_p^{\dagger}a_q^{\dagger}|0\rangle$,

\begin{equation}
\langle V_c \rangle_{\pm} =\pm 2|f(p,x)|^2|f(q,x)|^2\frac{1}{(p-q)^2},
\end{equation}
we find that the Coulomb energy is positive semi definite 
in the bosonic case, $\langle V_c \rangle_+\geq 0$,
and
negative semi definite in the fermionic case, $\langle V_c \rangle_-\leq 0$.
If we allow additional internal states with an identical momentum, i.e. 
$f_i(p,x)$ with $i=1,2$
and evaluate $\langle V_c\rangle$ 
with the use of a state, $a^{\dagger}_{p,i=1} a^{\dagger}_{p,i=2}|0\rangle $,
we find $\langle V_c\rangle=0$. Therefore, it turns out that
in the case of bosons, the state of two particles with an identical momentum
is more stable than the state of two particles with different momentum.
On the other hand, in the case of fermions, the situation is reverse;
the state of two particles with different momentum is more stable than the state
of two particles with an identical momentum.


To summarize, we have shown using light cone quantization that Savvidy instability is solved in
sufficiently dense quark matter where quarks and gluons couple weakly with each others.
In the quark matter the color magnetic field is generated spontaneously 
and gluons in the lowest Landau level are also produced spontaneously 
to form a ground state of gluons. 
All of gluons in the ground state carry a vanishingly small longitudinal momentum. Thus,
they form a nontrivial two dimensional state with finite color charge.
The state arises due to the effect of Coulomb interaction among gluons in the
longitudinal direction. We need further analysis to see whether or not these gluons form
quantum Hall states.

\vspace*{2em}
We would like to express thanks
to Prof. O. Morimatsu, Dr. T. Nishikawa
and Dr. M. Ohtani for useful discussion.
This work was supported by Grants-in-Aid of the Japanese Ministry
of Education, Science, Sports, Culture and Technology (No. 13135218).



\begin{thebibliography}{99}
\bibitem{savvidy}G.K. Savvidy, Phys. Lett. B 71 (1977) 133.
\bibitem{unstable}N.K. Nielsen and P. Olesen, Nucl. Phys. B 144 (1978) 376;
Phys. Lett. B 79 (1978) 304.
\bibitem{nielsen}J. Ambj\o{}rn, N.K. Nielsen and P. Olesen, Nucl. Phys. {\bf B152},75 (1979).\\

H.B. Nielsen and M. Ninomiya, Nucl. Phys. {\bf B156}, 1 (1979).\\
H. B. Nielsen and P. Olesen, Nucl. Phys. {\bf B160}, 330 (1979).
\bibitem{iwa}A. Iwazaki and O. Morimatsu, Phys. Lett. B 571 (2003) 61.\\ 
A. Iwazaki, O. Morimatsu, T. Nishikawa and M. Ohtani.\\
Phys. Lett. B 579 (2004) 347. A. Iwazaki, O. Morimatsu, T. Nishikawa and M. Ohtani,
Phys. Rev. D 71 (2005) 034014. 
\bibitem{qh}{\it The Quantum Hall Effect, 2nd Ed.}, edited by R.E. Prange
  and S.M. Girvan ( Springer-Verlag, New York, 1990 ).\\
S. Das Sarma and A. Pinczuk (Eds.),
Perspectives in Quantum Hall Effects, (Wiley, New York, 1997).
\bibitem{EI}Z.F. Ezawa, M. Hotta and A. Iwazaki, 
Phys. Rev. B46, 7765 (1992);
Z.F. Ezawa and A. Iwazaki, J. Phys. Soc. Jpn. 61, 4133 (1992).
\bibitem{color}K. Rajagopal and F. Wilczek, hep-ph/0011333.
\bibitem{magnetar}A. Iwazaki, Phys. Rev. D72 (2005) 114003.
\bibitem{cgc}A. Iwazaki, hep-ph/0604222.
\bibitem{light}P.A.M. Dirac, Rev. Mod. Phys. 21 (1949) 392.
\bibitem{cone}for a review, see A. Harindranath, hep-ph/9612244, 
T. Heintl, hep-th/0008096.
\bibitem{nakajima}T. Nakajima and M. Ueda, Phys. Rev. Lett. {\bf 91} 140401 (2003).
\bibitem{kondo}K. Kondo, Phys. Lett. B600, 287 (2004).
\bibitem{yamawaki}T. Maskawa and K. Yamawaki, Prog. Theor. Phys. 56 (1976) 270.
\bibitem{yamawaki2}S. Tsujimaru and K. Yamawaki, Phys. Rev. D57 (1998) 4942.

\bibitem{thorn}J.S. Rozowsky and C.B. Thorn, Phys. Rev. Lett. 85 (2000) 1614.
\bibitem{qcd}E. Iancu and R. Venugopalan, hep-ph/0303204.
\end{thebibliography}
\end{document}